\documentclass{webofc}
\usepackage[varg]{txfonts}
\usepackage{caption}
\usepackage{subcaption}
\begin{document}
\title{Oriented Event Shapes for massive Quarks}

\begin{flushright} IFT-UAM/CSIC-22-145\end{flushright}\vspace*{-2cm}

\author{\firstname{Alejandro} \lastname{Bris}\inst{1,2}\fnsep\thanks{\email{ alejandro.bris@uam.es}} \and
\firstname{N\'estor G.} \lastname{Gracia}\inst{3}\fnsep\thanks{\email{ngonzalez@usal.es}} \and
\firstname{Vicent} \lastname{Mateu}\inst{3}\fnsep\thanks{\email{vmateu@usal.es}}}

\institute{Departamento de F\'isica Te\'orica, Universidad Aut\'onoma de Madrid,\\E-28049, Madrid, Spain
\and Instituto de F\'isica Te\'orica UAM-CSIC,\\E-28049 Madrid, Spain
\and Departamento de F\'isica Fundamental e IUFFyM, Universidad de Salamanca,\\E-37008 Salamanca, Spain}

\abstract{In this work we present the computation of so-called oriented event-shape distributions for massive quarks
up to $\mathcal{O}(\alpha_s)$, along with the total oriented cross section in which one does not look at
the geometric properties of the momentum distribution for particles in the final state. We consider the
vector and axial-vector currents, and for the former, we find a non-vanishing result at $\mathcal{O}(\alpha_s^0)$
that translates into a significant enhancement as compared to the massless approximation. Our results are an
important ingredient for analyses that aim to determine the strong coupling with high accuracy.}
\maketitle
\section{Introduction}
\label{intro}
State-of-the art predictions for event shapes with massive quarks are less advanced than those for massless
particles. Similarly, our theoretical knowledge of oriented event shapes is significantly worse than that
of observables in which the orientation of the event is ignored. We aim to reduce these two gaps by computing
the NLO fixed-order expressions of massive-quark-initiated oriented event shapes, for differential, cumulative
and total cross sections.

We consider the process $e^+e^-\to$ hadrons initiated by a $Q\overline Q$ pair, with $Q$ a massive quark, and
define the event orientation by the angle $\theta_T$ between the thrust axis and the beam direction. The thrust
axis ${\hat n}$ is the unit vector appearing in the definition of thrust~\cite{Farhi:1977sg}:
\begin{equation}\label{eq:axisDef}
\tau = 1 - \max_{\hat n}\frac{\sum_i |\vec{p}_i\!\cdot \hat{n}|}{\sum_i |\vec{p}_i|}\,,
\end{equation}
that maximizes the sum. It can be shown that it is always parallel to the sum of the \mbox{$3$-momenta} of a subset of final-state particles
within the same hemisphere.
In Ref.~\cite{Mateu:2013gya} it was shown that to fully determine the orientation of the cross section it is enough to specify two
$\theta_T$-independent structures:
\begin{equation}
\frac{1}{\sigma_0} \frac{{\rm d} \sigma}{{\rm d} \!\cos (\theta_T) {\rm d} e} =\frac{3}{8}[1+\cos^2(\theta_T)]
\frac{1}{\sigma_0} \frac{{\rm d} \sigma}{{\rm d} e}
+[1 - 3 \cos^2(\theta_T)]\frac{1}{\sigma_0} \frac{{\rm d} \sigma_{\rm ang}}{{\rm d} e}\,,
\end{equation}
with $e$ a generic event shape and $\sigma_0$ the Born cross section. The first term is the unoriented
distribution, while the second, dubbed the angular term, is the one we focus on. In the rest of this
write-up we denote the minimal value that the event shape $e$ can take by $e_{\rm min}$. It does not depend
on the number of partons in the final state, and it is attained in the physical situation in which the quark
and the anti-quark are produced in conjunction with any number of massless particles with zero energy. Therefore
the soft singularities will take place for $e\to e_{\rm min}$, that is, for dijet configurations.

In our computations we use dimensional regularization to deal with both ultraviolet (UV) and infrared (IR)
divergences. For the latter one has to work out the $2$- and $3$-particle phase space in $d=4-2\varepsilon$
dimensions, differential in the polar angles of the quarks. To that end we used the Gram-Schmidt procedure
to consistently construct a suitable set of axes in our vector space with a non-integer number of dimensions.
When combining real- and virtual-radiation contributions we end up with a finite result.

\section{Lowest order Result}
\begin{figure*}[t!]\centering
\includegraphics[width=0.4\textwidth]{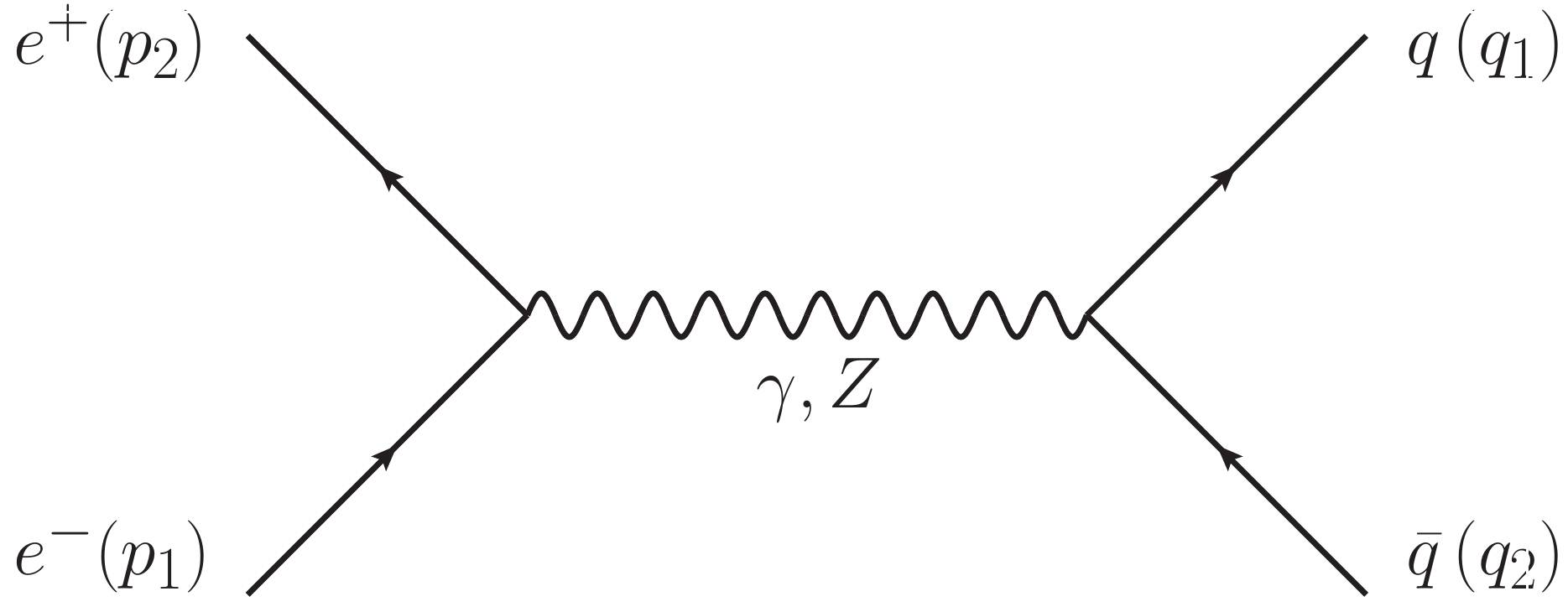}
\includegraphics[width=0.21\textwidth]{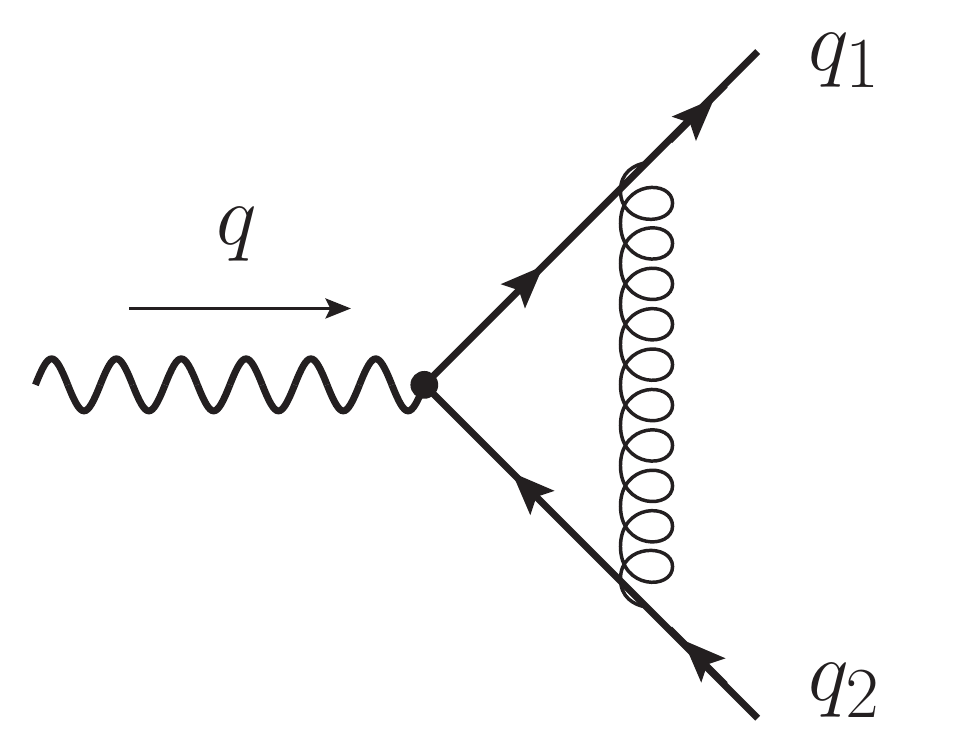}
\caption{Feynman diagrams for the partonic process $e^+e^-\to Q\overline Q$ at LO and NLO.}
\label{fig:TreeVirt}
\end{figure*}
We start off with the $\mathcal{O}(\alpha_s^0)$ results, which are computed in $d=4-2\varepsilon$ dimensions
even though they are UV and IR finite. This is still very useful as one can normalize the cross sections with
the $d$-dimensional point-like cross section to have nice-looking intermediate results. The relevant Feynman
diagram is shown in the left panel of Fig.~\ref{fig:TreeVirt}. At this point it is
instructive to introduce the total angular cross section:
\begin{equation}
R_{\rm ang} = \int {\rm d}e \frac{1}{\sigma_0} \frac{{\rm d} \sigma_{\rm ang}}{{\rm d} e}
\equiv \sum_{n=0}^\infty \biggl[\frac{\alpha_s(\mu)}{\pi}\biggr]^{n} R_n^{{\rm ang},C},
\end{equation}
which does not depend on the specific choice for $e$. Using the $d$-dimensional $2$-particle phase space
differential in the polar angle
\begin{equation}\label{eq:2body}
\frac{1}{2s} \frac{{\rm d} Q_2}{{\rm d}\! \cos (\theta)} = \frac{\beta^{1 - 2 \varepsilon} \sin^{- 2 \varepsilon} (\theta)}{2^{5 - 4
\varepsilon} s^{1 + \varepsilon} \Gamma (1 - \varepsilon) \pi^{1 -
\varepsilon}}\,,
\end{equation}
with $s=Q^2$ the center-of-mass energy squared, and where we have already included the flux factor, a relatively
simple computation yields~\cite{Bris:2022cdr}\footnote{All
results are presented in the pole mass scheme.}
\begin{equation}
\sigma_B = N_c Q_q^2 \frac{(4 \pi)^{1 + \varepsilon} (1 - \varepsilon) \Gamma (2 -
\varepsilon) \alpha_{\rm em}^2}{(3 - 2 \varepsilon)
\Gamma (2 - 2 \varepsilon) s^{1 + \varepsilon}},\qquad
R_0^{{\rm ang},V} = \frac{3 \hat m^2\beta}{4}\,,\qquad
R_0^{{\rm ang},A} = 0\,,
\end{equation}
where we have set $\varepsilon=0$ in $R_0^{{\rm ang},C}$ but kept an arbitrary $d$ in the point-like
cross section. Here $\beta\equiv\sqrt{1-4\hat m^2}$ is the quark velocity and $\hat m = m/Q$ its
reduced mass. The non-vanishing result for $R_0^{{\rm ang},V}$ implies an enhancement with respect
to the massless approximation and the appearance of IR divergences at NLO, which translate into singular
distributional structures in differential cross sections. These results can be converted into differential
distributions simply multiplying by $\delta(e-e_{\rm min})$.
The LO result for $R_0^{{\rm ang},V}$ is shown graphically in the left panel of
Fig.~\ref{fig:R-tree-loop}, together with its SCET and threshold approximations.

\section{Virtual radiation}
\begin{figure*}[t!]\centering
\includegraphics[width=0.43\textwidth]{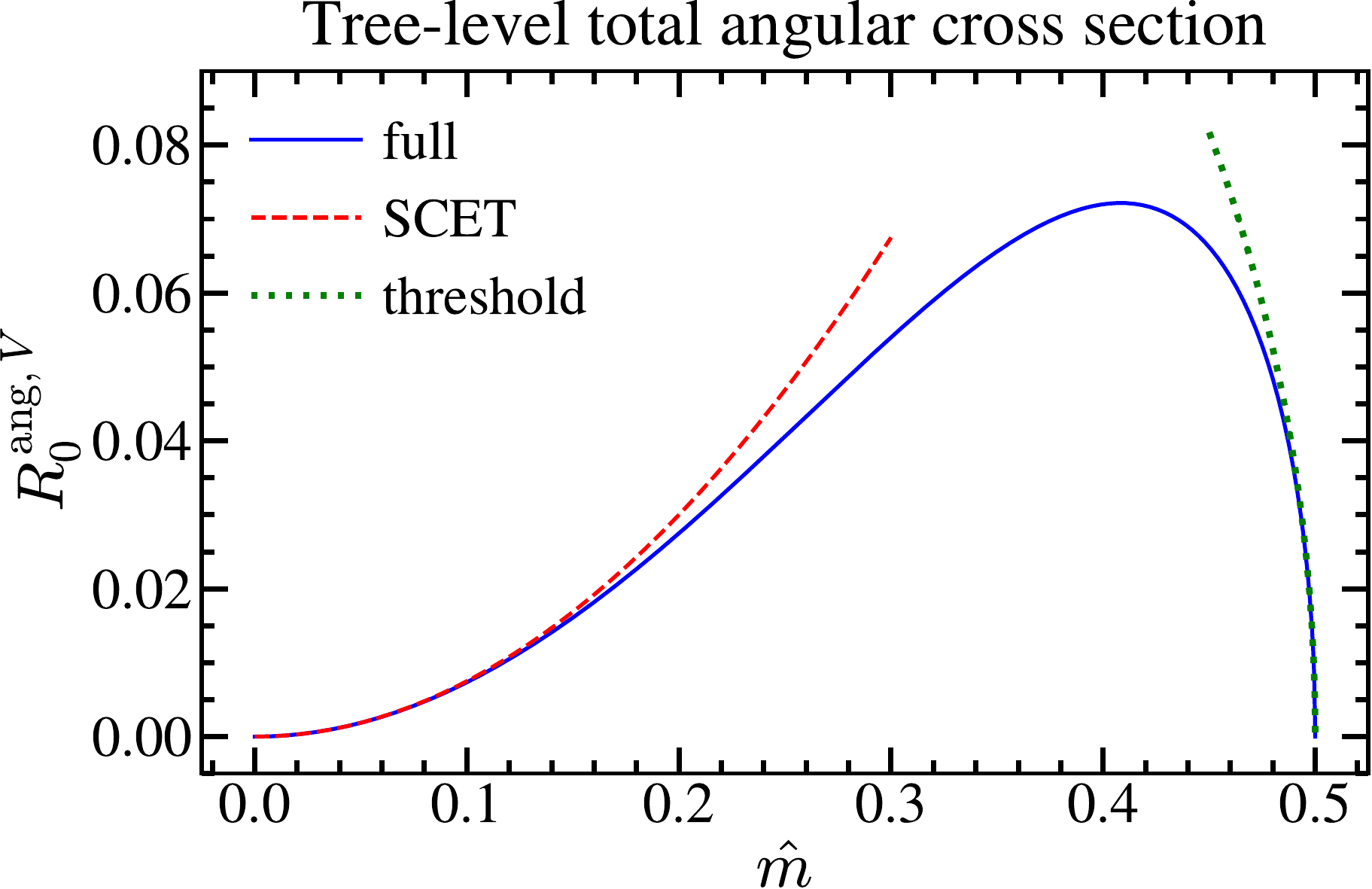}~~~~~~~~
\includegraphics[width=0.42\textwidth]{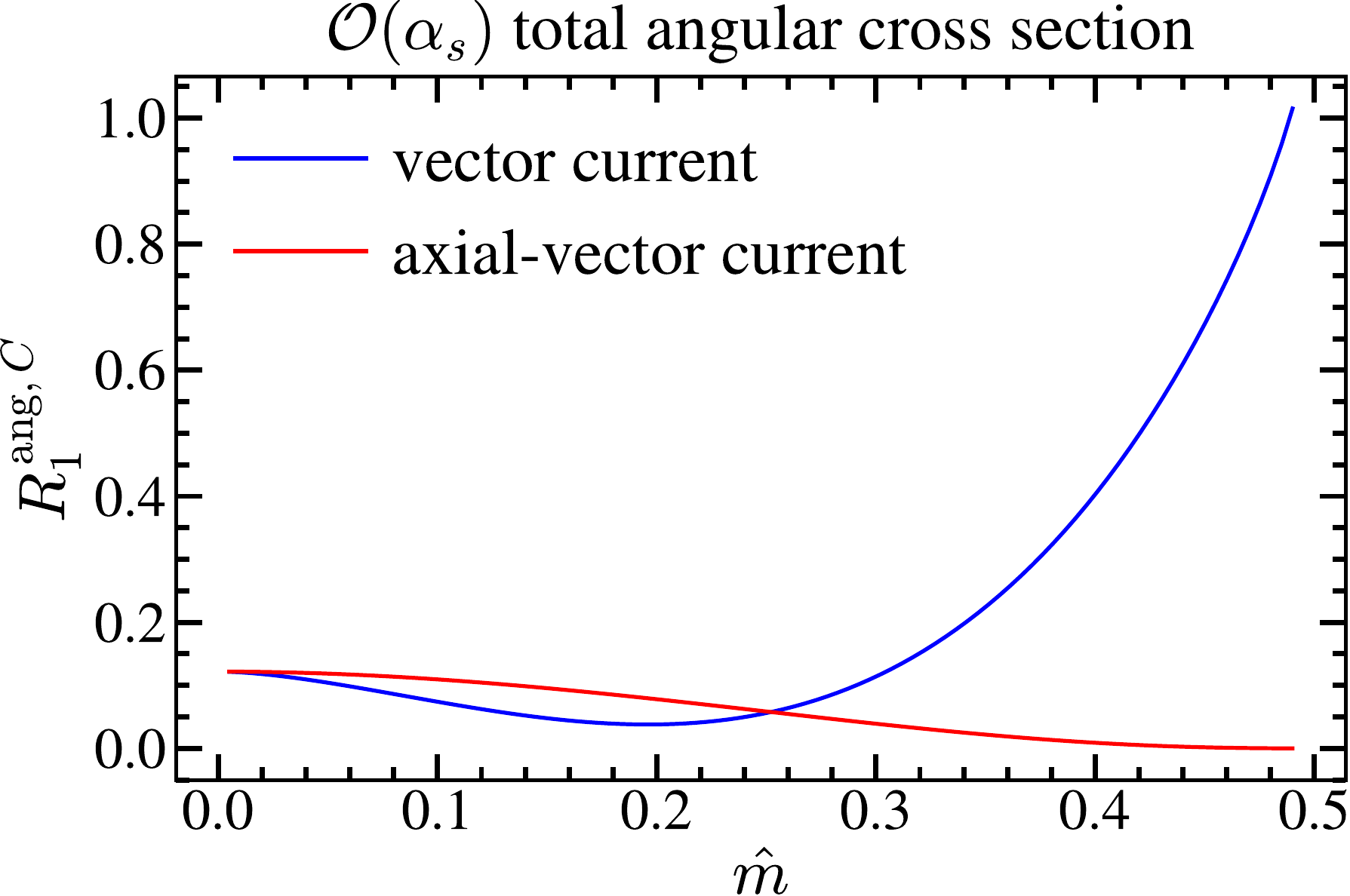}
\caption{Total angular cross section at LO for the vector current (left panel) and NLO (right panel)
for vector (blue) and axial-vector (red) currents.}
\label{fig:R-tree-loop}
\end{figure*}
The relevant diagram for this contribution appears in the right panel of Fig.~\ref{fig:R-tree-loop}.
To carry out this computation we use the known result for the massive quark form factors for vector and
axial-vector currents
\begin{align}
V^{\mu} & = \biggl[ 1 + C_F \frac{\alpha_s}{\pi} A (\hat{m}) \biggr]
\gamma^{\mu} + C_F \frac{\alpha_s}{\pi} \frac{B (\hat{m})}{2 m} (q_1 - q_2)^{\mu} , \\
A^{\mu} & = \biggl[ 1 + C_F \frac{\alpha_s}{\pi} C (\hat{m}) \biggr]
\gamma^{\mu} \gamma_5 + C_F \frac{\alpha_s}{\pi} \frac{D (\hat{m})}{2 m}\, \gamma_5\, q^{\mu}, \nonumber
\end{align}
with $q = q_1 + q_2$ and $q_{1,2}$ the photon, quark and anti-quark momenta, respectively. Only the terms
$A, B$ and $C$ contribute due to vector current conservation on either the quark or lepton sides, and we
only need the real part of them~\cite{Jersak:1981sp, Harris:2001sx}
\begin{align}
{\rm Re} [A (\hat{m})] &= \biggl( \frac{1+\beta^2}{2\beta}\log\biggl(\frac{1 + \beta}{2 \hat{m}}\biggr)
-\frac{1}{2} \biggr)
\biggl[\frac{1}{\varepsilon} - 2\log\biggl( \frac{m}{\mu} \biggr)\biggr]+
A_{\rm reg}(\hat{m})\,,\\
A_{\rm reg}(\hat{m}) & =\frac{3}{2} \beta \log\biggl(\frac{1 + \beta}{2 \hat{m}}\biggr) - 1 + \, \frac{1+\beta^2}{4\beta} \biggl[ \pi^2 - 2
\log^2\biggl[\frac{1 + \beta}{2 \hat{m}}\biggr] - 2\, {\rm Li}_2\biggl( \frac{2 \beta}{1 + \beta} \biggr) \biggr]\nonumber\,,\\
{\rm Re} [C (\hat{m})] & = {\rm Re} [A (\hat{m})] + \frac{4\hat{m}^2 }{\beta}
\log\biggl(\frac{1 + \beta}{2 \hat{m}}\biggr)\,.\nonumber
\end{align}
Keeping an arbitrary $d$ in both currents, after adding the flux factor and integrating the \mbox{$d$-dimensional} $2$-body phase space
we find~\cite{Bris:2022cdr}
\begin{align}\label{eq:virtual}
R_{21}^{{\rm ang},V} \,& = \frac{\beta}{2}\biggl\{ \biggl[ 1 - \frac{2 }{\beta}(1 - 2 \hat{m}^2)
\log\biggl(\frac{1 + \beta}{2 \hat{m}}\biggr)
\biggr] \biggl[ 2 \hat{m}^2 \log \biggl(
\frac{m}{\mu} \biggr) - \frac{\hat{m}^2}{\varepsilon} + \frac{3}{10}(3 - 2 \hat{m}^2) + 2 \hat{m}^2 \log
(\beta) \biggr] \nonumber\\
& + 2 \hat{m}^2 {\rm Re} [A_{\rm reg} (\hat{m})] - \hat{m}^2 \beta
\log\biggl(\frac{1 + \beta}{2 \hat{m}}\biggr) \biggr\}\, ,\nonumber\\
R_{21}^{{\rm ang},A} \,& = \frac{9 \beta^2}{5} \biggl[ \frac{\beta}{2} -
(1 - 2 \hat{m}^2)\log\biggl(\frac{1 + \beta}{2 \hat{m}}\biggr) \biggr]\,,
\end{align}
where the $1/\varepsilon$ singularity in the vector current is of IR origin. To convert these two
results into differential distributions one simply multiplies them by $\delta(e-e_{\rm min})$.
The fact that our results do not vanish in the massless limit is an artifact of dimensional
regularization. When added to the real-radiation contribution all artifacts and divergences cancel.

\section{Real radiation}
\begin{figure}[t]\centering
\includegraphics[width=0.75\textwidth]{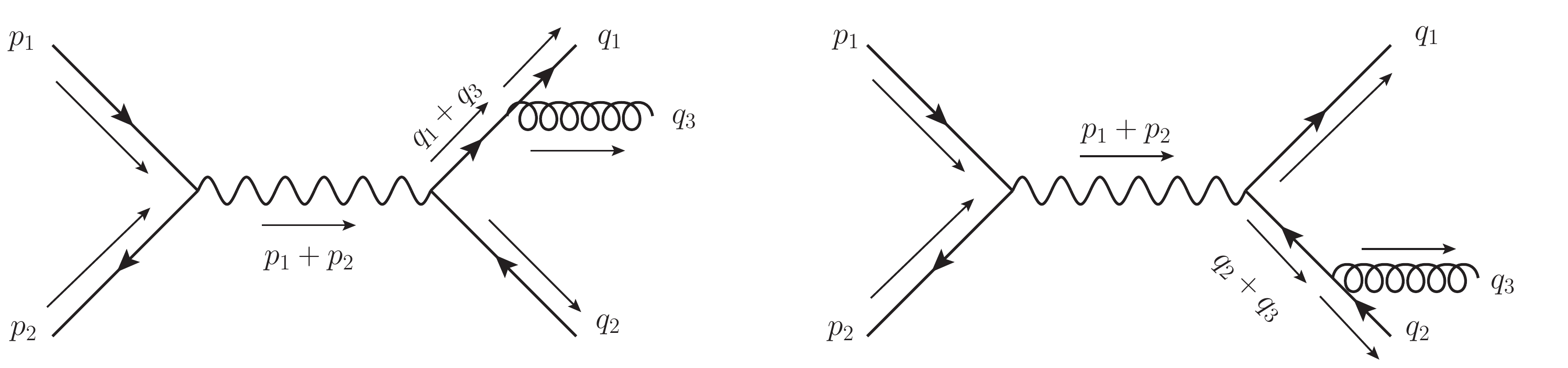}
\caption{Diagrams for NLO real-radiation contributions.\label{fig:real}}
\end{figure}
Before discussing the real-radiation computation, we show the general form of the
differential distribution for massive quarks up to NLO:
\begin{align}\label{eq:general-diff}
\frac{1}{\sigma^C_0} \frac{{\rm d} \sigma^C_{\rm ang}}{{\rm d} e} =\,&
R^{0,C}_{\rm ang}(\hat m)\,\delta [e - e_{\rm min}(\hat m)] +
C_F \frac{\alpha_s(\mu)}{\pi} A^{{\rm ang},C}_{e}({\hat m})\delta [e - e_{\rm min}(\hat m)] \\
& + C_F \frac{\alpha_s(\mu)}{\pi} B^{{\rm ang},C}_{\rm plus}({\hat m})
\biggl[\frac{1}{e - e_{\rm min}(\hat m)} \biggr]_+
+ C_F \frac{\alpha_s(\mu)}{\pi} F^{\rm ang}_{C,e} (e, \hat{m}) + \mathcal{O}(\alpha_s^2)\,,\nonumber
\end{align}
where $F^{\rm ang}_{C,e}$ is an integrable function in the vicinity of $e=e_{\rm min}$ and the rest of structures
are distributions globally referred to as `singular terms'. It is easy to see that $F^{\rm ang}_{C,e}$ and
$B^{{\rm ang},C}_{\rm plus}$ come only from diagrams with more than two particles in the final state while
$A^{{\rm ang},C}_{e}$ has contributions from both virtual and real radiation. Finally, for the axial-vector current,
all terms except for $F^{\rm ang}_{A,e}$ are zero.

The modulus squared of the sum of amplitudes shown in Fig.~\ref{fig:real}, sumed over final polarizations and
averaged over initial ones, can be written as
\begin{equation}\label{eq:general4}
\overline\sum_\lambda|M|^2=\frac{8 \pi^2 C_F}{s}\biggl(\frac{\mu^2e^{\gamma_E}}{4\pi}\biggr)^{\!\varepsilon}
[A^C_0 + A^C_1 \beta_1^2 \cos^2 (\theta_1) + A^C_2 \beta_2^2 \cos^2 (\theta_2)
+ A^C_{12} \beta_1 \beta_2 \cos (\theta_1) \cos (\theta_1)]\,,
\end{equation}
where $\theta_{1,2}$ are the angles between the beam and the quark/anti-quark $3$-momentum. The analytic form
of the $A_i^C$ coefficients will be given elsewhere. To translate this result into a cross section we need
the $d$-dimensional $3$-particle phase space, differential in the polar angles of the quark and the anti-quark:
\begin{align}\label{eq:Q3}
\frac{{\rm d} Q_3}{2 s} =\,& \frac{4^{\varepsilon} s^{- 2 \varepsilon}}{2
(4 \pi)^{4 - 2 \varepsilon} \Gamma (1 - 2 \varepsilon)} \! \int \!
{\rm d} x_1 {\rm d} x_2 {\rm d} \!\cos (\theta_i) {\rm d}\! \cos (\theta_j)
\frac{\beta_i^{-2\varepsilon}\beta_j^{-2\varepsilon}\theta(h_{i j})}{h_{i j}^{1 / 2 + \varepsilon}}\,,\\
h_{ij} = \,& \sin^2 (\tilde{\theta}_{ij}) - \cos^2 (\theta_i) - \cos^2 (\theta_j) + 2 \cos
(\tilde{\theta}_{ij}) \cos (\theta_i) \cos (\theta_j)\nonumber\\
\equiv\,& [\cos (\theta_i) - \cos (\theta_{ij}^-)] [\cos (\theta_{ij}^+) - \cos(\theta_i)]\,,\nonumber\\
\cos (\theta_{ij}^{\pm}) = \,&\cos (\tilde{\theta}_{ij}) \cos (\theta_j) \pm \sin (\tilde{\theta}_{ij})\sin (\theta_j)
= \cos (\tilde{\theta}_{ij} \mp \theta_j)\,.\nonumber
\end{align}
where $x_i=2E_i/Q$ are dimensionless variables proportional to the energy of each particle. We use $i=1,2$ and $3$ to
label the quark, antiquark and gluon, respectively. We have defined $\beta_i\equiv2|\vec{p}_i|/Q=\sqrt{x_i^2-4\hat m_i^2}$.
The results above serve to compute the multi-differential cross section at NLO:
\begin{align}
\frac{1}{\sigma^C_0} \frac{{\rm d}^4 \sigma^C_{\alpha_s}}{{\rm d} x_1 {\rm d} x_2 {\rm d}\! \cos
(\theta_i) {\rm d}\! \cos (\theta_j)} =\, & \frac{4^{\varepsilon} \alpha_s
C_F}{16 \pi^2} \frac{(3 - 2 \varepsilon) (1 - 2 \varepsilon)}{(1 -
\varepsilon) \Gamma (2 - \varepsilon)} \biggl( \frac{\mu^2e^{\gamma_E}}{s}
\biggr)^{\!\!\varepsilon} \frac{\beta_i^{-2\varepsilon} \beta_j^{-2\varepsilon}}{h_{i j}^{1 / 2 + \varepsilon}} \\
& \times\![A^C_0 + A^C_1 \beta_1^2 \cos^2 (\theta_1) + A^C_2 \beta_2^2 \cos^2 (\theta_2)\nonumber\\
& \quad + A^C_{12} \beta_1 \beta_2 \cos (\theta_1) \cos (\theta_1)]\,,
\nonumber
\end{align}
From the expression above we can project out the angular
distribution differential in the quark and anti-quark energy in $d$ dimensions:
\begin{align}\label{eq:realProj}
& \frac{1}{\sigma_0^C} \frac{{\rm d}^2 \sigma^{\alpha_s,C}_{\rm ang}}{{\rm d} z {\rm d} y}
= \frac{3 \alpha_s C_F}{8 \pi} \frac{y^{1 - 2 \varepsilon}}{(1 -
\varepsilon)^2 \Gamma (1 - \varepsilon)} \biggl( \frac{\mu^2e^{\gamma_E}}{s}
\biggr)^{\!\varepsilon}[ (1 - y) (1 - z) z \, - \hat{m}^2 ]^{-
\varepsilon} \\
&\qquad\qquad\quad\times\! \biggl\{ A^C_q(\hat m, y, z) \theta \biggl(z-\frac{1}{2}\biggr) \theta [y_{\tau} (\hat{m}, 1-z) - y ]
+ A^C_{\bar q}(\hat m, y, z) \theta \biggl(\frac{1}{2}-z\biggr) \theta [y_{\tau} (\hat{m}, z) - y ]\nonumber\\
&\qquad\qquad\qquad+ A^C_g(\hat m, y, z) \theta [y - y_{\tau} (\hat{m}, z) ] \theta [y - y_{\tau} (\hat{m}, 1-z)] \biggr\},
\nonumber
\end{align}
\begin{figure}[t]\centering
\includegraphics[width=0.4\textwidth]{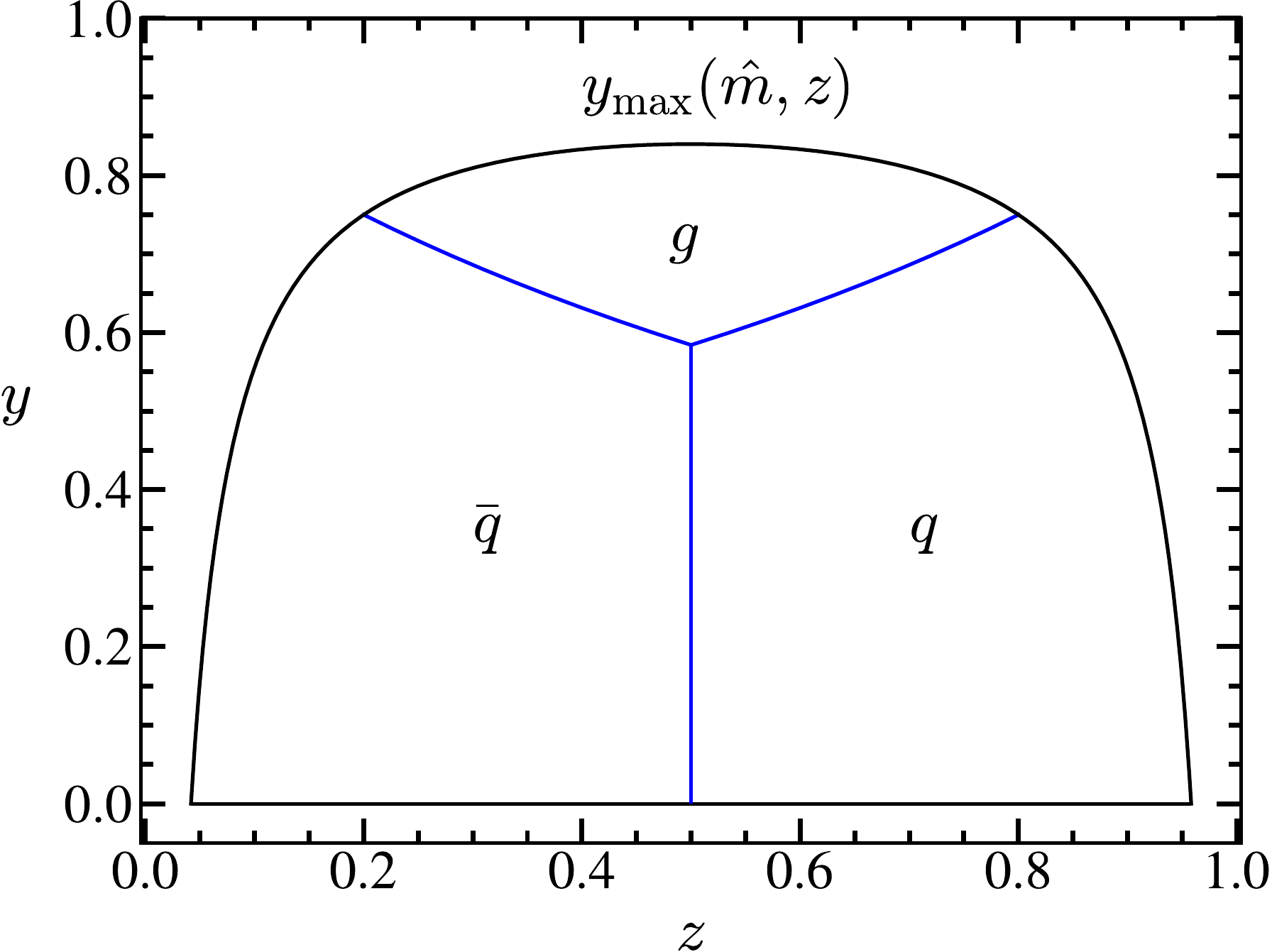}
\caption{Phase space for three particles (two quarks with equal mass plus a massless gluon) in $(z,y)$ coordinates.
\label{fig:Phase-space}}
\end{figure}
\!\!where we used the symmetric kinematic variables $(y,z)$ defined through the change of variables
\mbox{$x_1 = 1 - (1 - z) y$}, $x_2 = 1 - z y$, such that the soft limit has been mapped to $y\to 0$.
The form of the phase space in these variables is shown in Fig.~\ref{fig:Phase-space}.
The functions $A_{q,\bar q, g}^C$ depend linearly on the $A_i^C$ defined in Eq.~\eqref{eq:general4}
with coefficients that depend on $y,z,\hat m$ and $\varepsilon$. The exact relation can be found in Ref.~\cite{Bris:2022cdr}, but
we provide the result after working out the combination for the axial current setting $\varepsilon=0$ ---\,since there
are no IR singularities\,--- [we do not show results for $A_{\bar q}^C$ since by symmetry arguments one can show that
$A_{\bar q}^C(\hat m,y,z)=A_q^C(\hat m,y,1-z)$]
\begin{align}
A_q^A(\hat m,y,z) = \,& \frac{(1 - y) z^2 - \hat{m}^2 z \{ 2 - y^2 + z [2
+ y (y - 2)] \} \!+ 2 \hat{m}^4}{z^2 \{ [1 - y (1 - z)]^2 - 4 \hat{m}^2 \}},\\
A_g^A(\hat m,y,z) = \,&\frac{2 (1 - y) (1 - z)^2 z^2 - \hat{m}^2 (1 - z)
z (4 - y^2 - 2 y) + 2 \hat{m}^4}{y^2 (1 - z)^2 z^2} . \nonumber
\end{align}
For the vector current, keeping only the linear dependence in $\varepsilon$ where it is strictly necessary we find~\cite{Bris:2022cdr}
\begin{figure*}[t!]\centering
\includegraphics[width=0.46\textwidth]{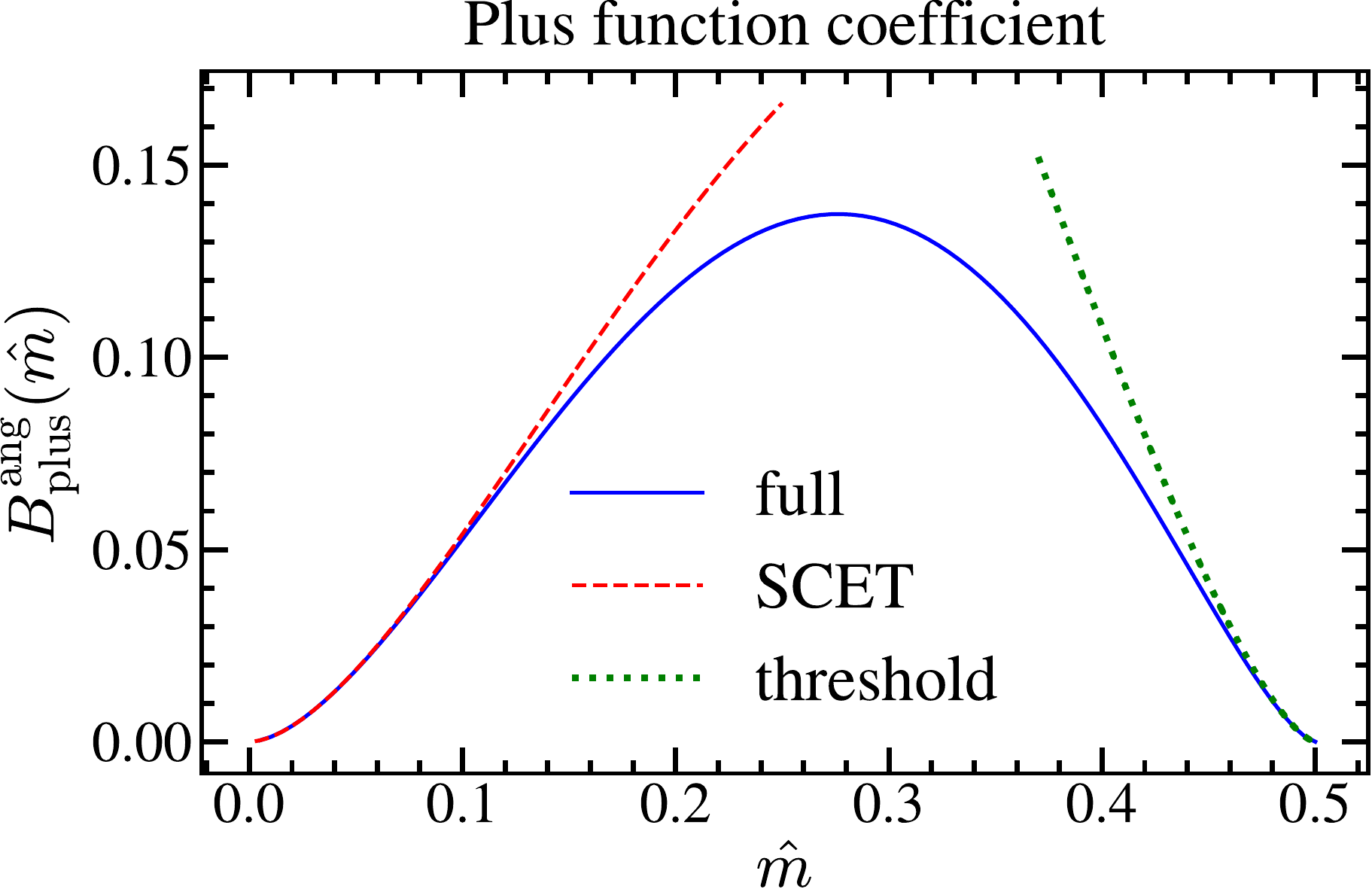}~~~~~
\includegraphics[width=0.45\textwidth]{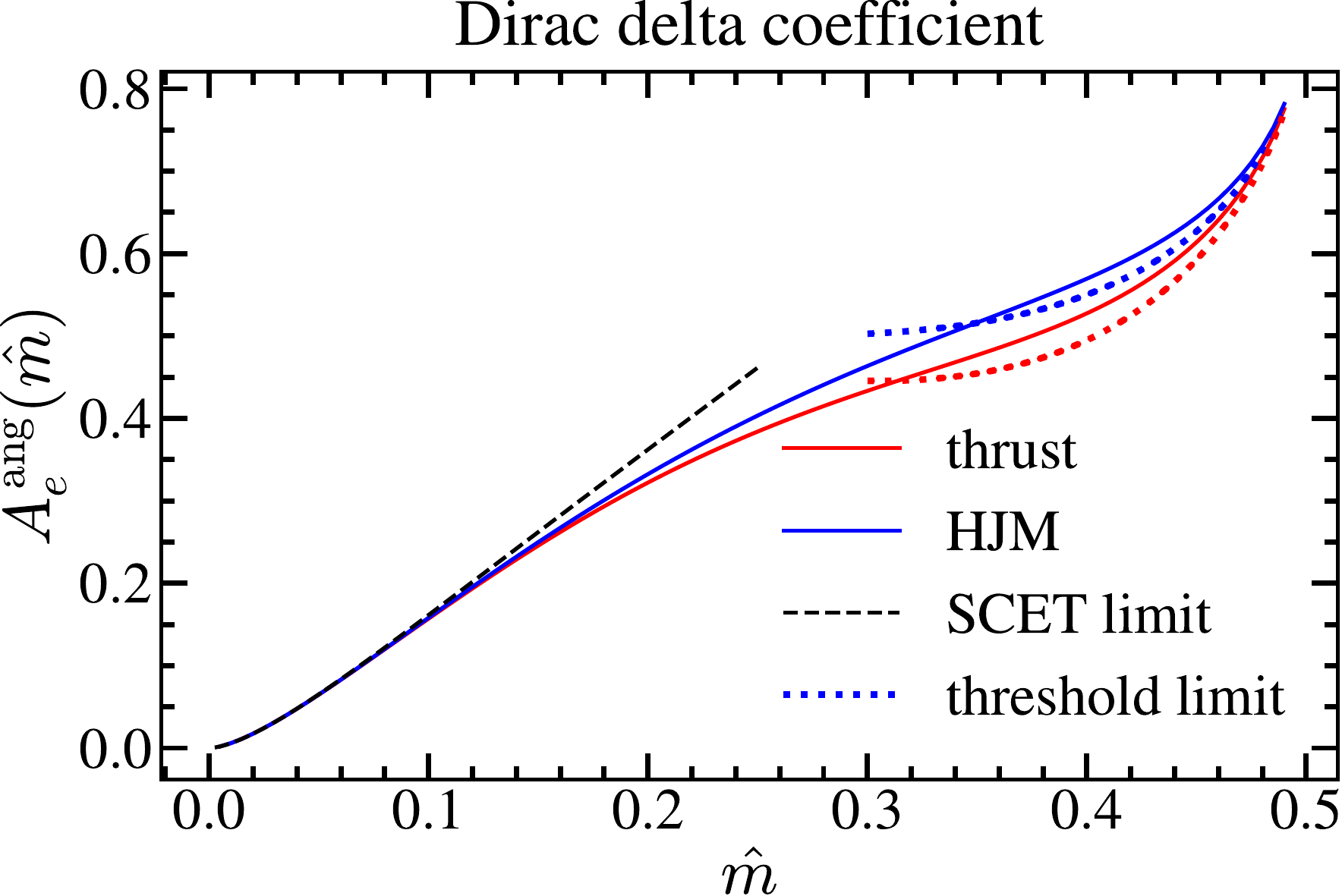}
\caption{Coefficients of the singular structures for the vector-current differential distribution,
as appearing in Eq.~\eqref{eq:singRes}.}
\label{fig:DeltaPlus}
\end{figure*}
\begin{align}
A_q^V(\hat{m},y,z) =\, & \frac{V^{\rm div} (\hat{m}, z) +
V^{\varepsilon} (\hat{m},z) \varepsilon}{y^2} + V^{\rm fin} (y, z,
\hat{m})\,, \\
V^{\rm fin} (\hat{m},y, z) =\, & \frac{(1 - y) y (1 - z) z^2 - \hat{m}^2
z [2 - y (1 - 2 z^2)] + 2 \hat{m}^4 [y (1 - z) + 4 z]}{y (1 - z) z^2 \{ [1 -
y (1 - z)]^2 - 4 \hat{m}^2 \}} \,. \nonumber\\
V^{\rm div} (\hat{m},z) = & - \!2 \hat{m}^2 M_V^1 (\hat{m},z)\,,
\qquad\qquad\quad~\, V^{\varepsilon} (\hat{m},z) =
\frac{9 + 14 \hat{m}^2}{5} M_V^1 (\hat{m},z)\,,\nonumber\\
A_g^V(\hat{m},y,z) = \,& \frac{2 [(1 - y) z (1 - z) - \hat{m}^2]}{y^2 (1 -
z) z}, \qquad
M_V^1 (\hat{m},z) = -\! \frac{(1 - z) z - \hat{m}^2}{(1 - z)^2 z^2}\,.\nonumber
\end{align}
In terms of those we find the following analytic results for the vector-current coefficients multiplying the
singular structures~\cite{Bris:2022cdr}:
\begin{align}\label{eq:singRes}
A^{\rm ang}_{e}({\hat m}) = \,& \frac{3}{4}[A_\delta(\hat m) - 2 \hat{m}^2 I_e (\hat{m})]\,,\qquad\quad
B^{\rm ang}_{\rm plus}({\hat m}) = \frac{3\hat m^2}{2} \biggl[2 (1 - 2 \hat{m}^2)
\log\biggl(\frac{1 + \beta}{2 \hat{m}}\biggr) - \beta\biggr] \,,\\
A_{\delta} (\hat{m}) = \, &\hat{m}^2 \biggl\{ 2 \beta [\log
(\hat{m}) - 1] + \frac{1 + \beta^2}{2} \biggl[ \pi^2 - 2 L^2_{\beta} +
{\rm Li}_2 \biggl( \frac{2 \beta}{\beta - 1} \biggr)
- 3\, {\rm Li}_2 \biggl( \frac{2 \beta}{\beta + 1} \biggr) \nonumber\\
&\qquad -\! 4 \log\biggl(\frac{1 + \beta}{2 \hat{m}}\biggr) [\log (\hat{m}) - 1] \biggr] \biggr\} ,\nonumber\\
I_e (\hat{m}) =\, &\! -\!\! \int_{z_-}^{\frac{1}{2}} {\rm d} z M_V^1 (\hat{m},z) \log [f_e (z)]\,,\nonumber
\end{align}
where $I_e(\hat m)$ has been analytically computed in Ref.~\cite{Lepenik:2019jjk} for a large set of event shapes. We present
graphically the form of $A^{\rm ang}_{e}({\hat m})$ for a couple of event shapes in Fig.~\ref{fig:DeltaPlus},
along with the universal $B^{\rm ang}_{\rm plus}$ coefficient.

\section{Numerical Analysis}
\begin{figure*}[t!]\centering
\includegraphics[width=0.4\textwidth]{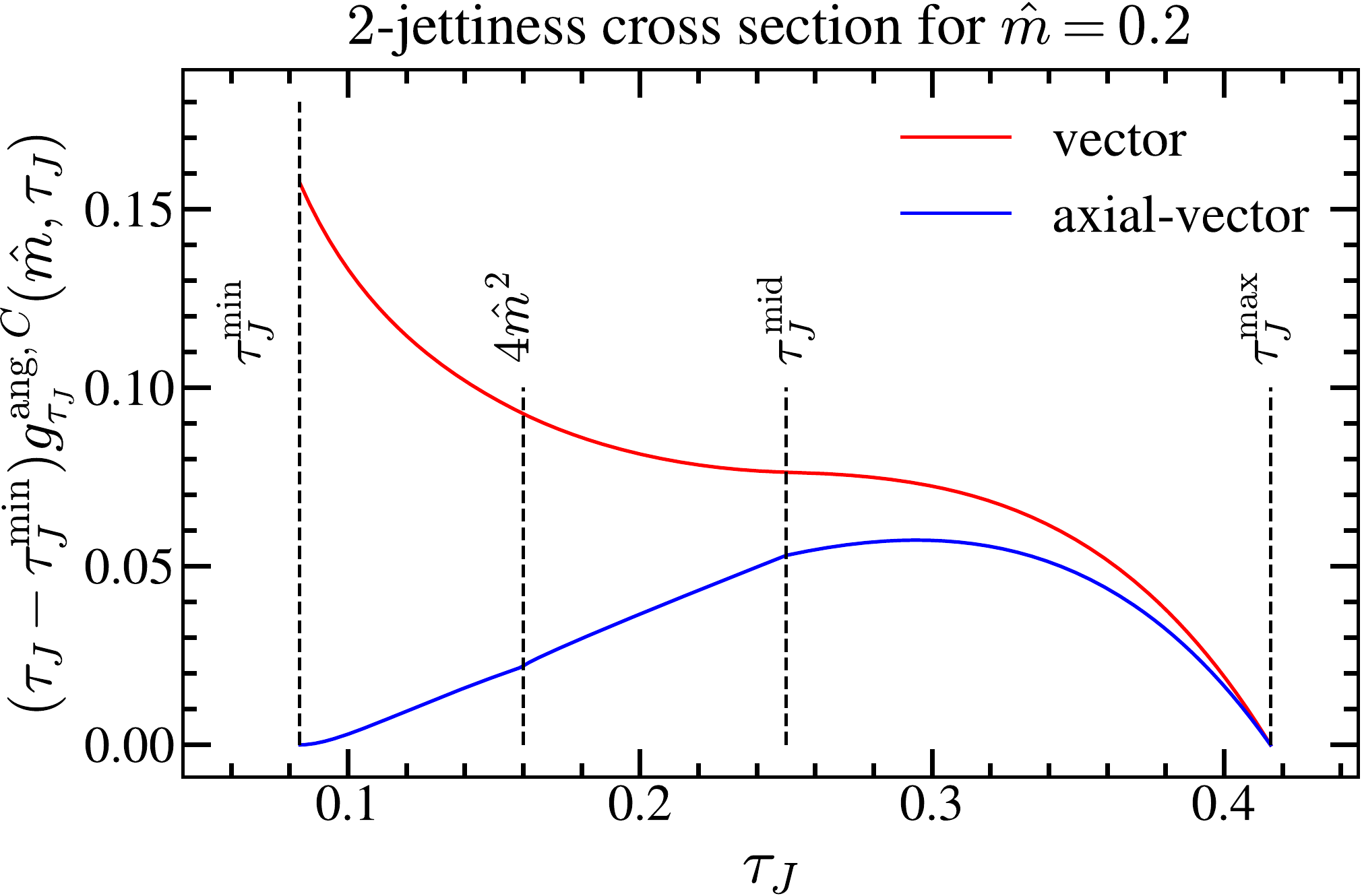}~~~~
\includegraphics[width=0.4\textwidth]{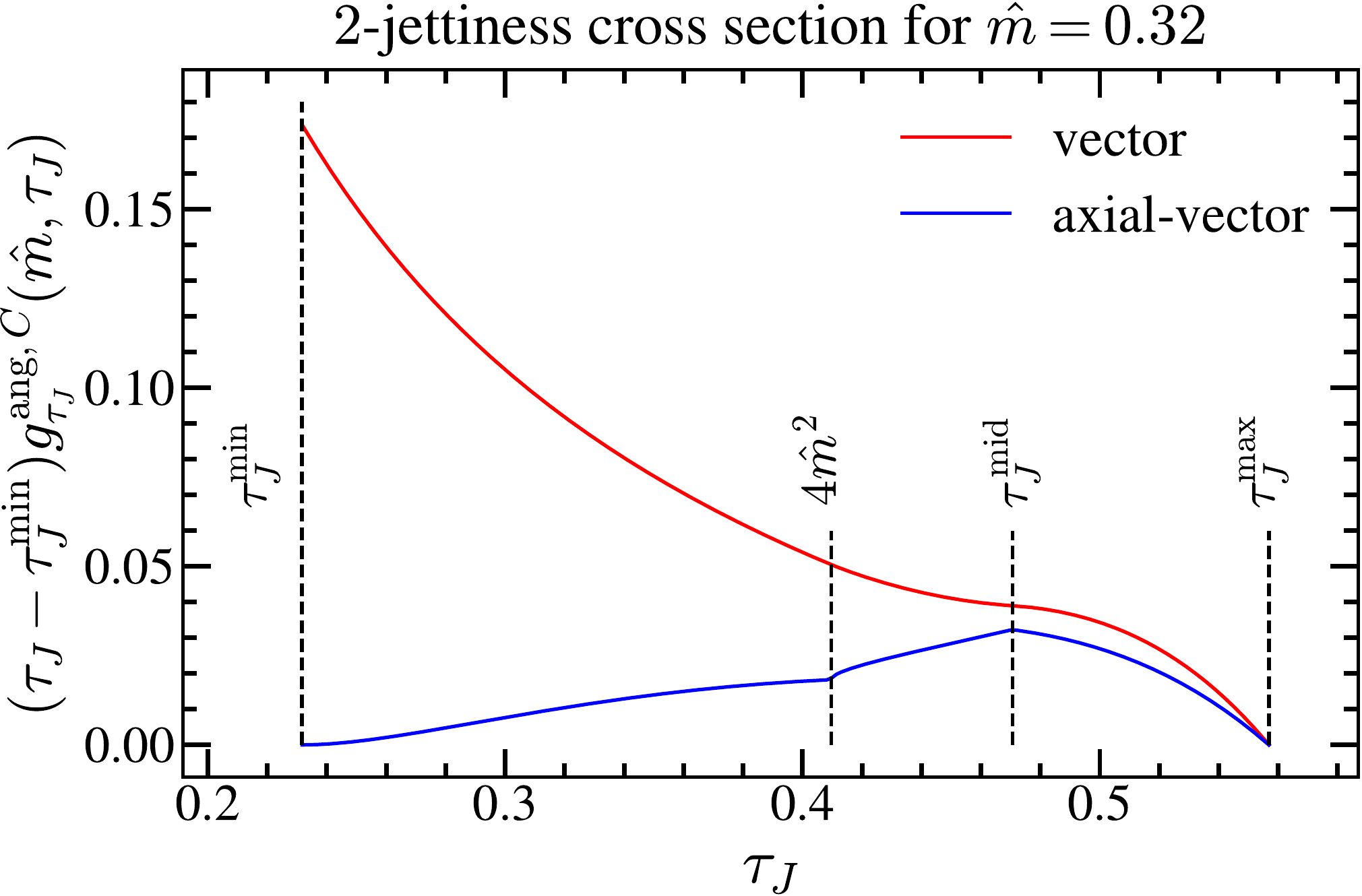}
\includegraphics[width=0.41\textwidth]{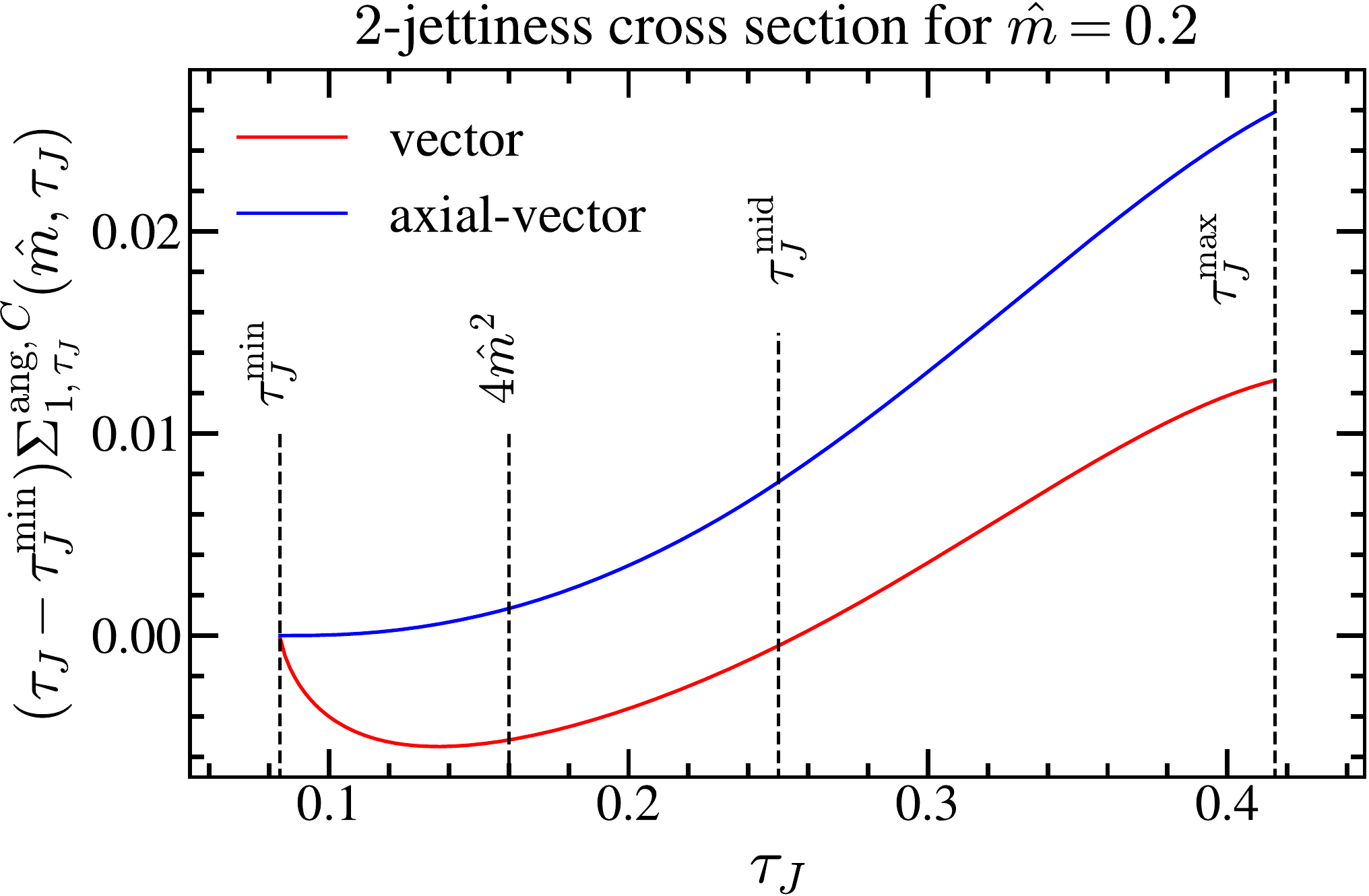}~~~~
\includegraphics[width=0.41\textwidth]{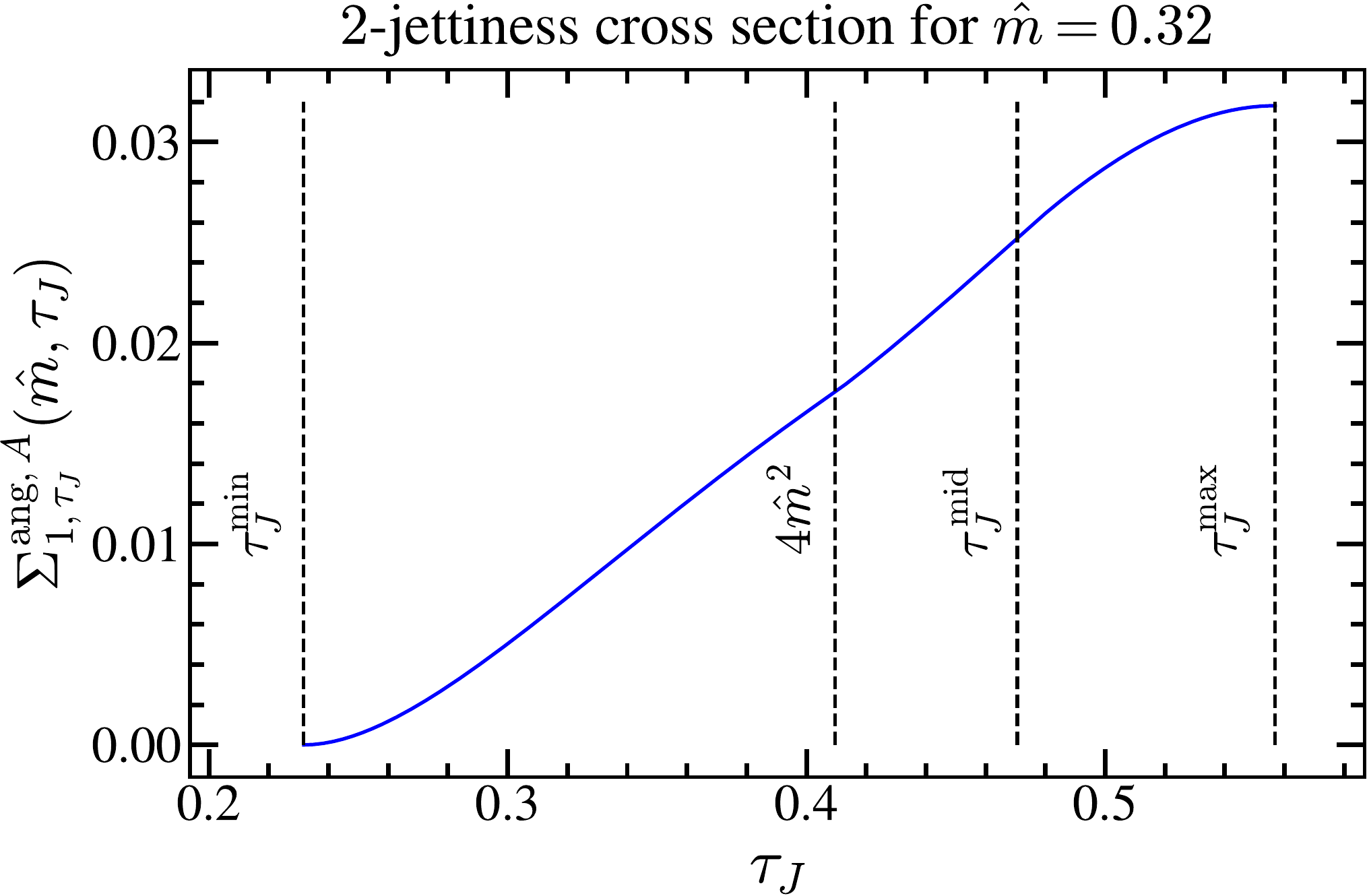}
\caption{Thrust differential and cumulative cross sections for values of the reduced
mass below and above $m_{\rm crit}$.} \label{fig:tau-dif-cum}
\end{figure*}

Integrating the differential distribution of Eq.~\eqref{eq:realProj} in the full $d$-dimensional phase space and
adding the virtual-diagram contribution one obtains the NLO total angular cross section. Explicit analytic
results can be found in Ref.~\cite{Bris:2022cdr}, but a graphical representation is shown in the right panel of
Fig.~\ref{fig:R-tree-loop}. The computation of the differential and cumulative distributions for $e>e_{\rm min}$
can be organized such that only $d=4$ expressions become necessary. For the mass-sensitive
event shapes $2$-jettiness~\cite{Stewart:2009yx} (closely related to thrust) and heavy jet
mass~\cite{Clavelli:1979md, Chandramohan:1980ry, Clavelli:1981yh} we have found
analytic expressions for
the differential distributions, while we have been able to write down their cumulative counterparts in terms of
1-dimensional numerical integrals. We do not show explicit expressions here, but present results in a numerical
form in Fig.~\ref{fig:tau-dif-cum}, where both vector and axial-vector currents are represented for two values
of the reduced mass $\hat m$. We mark with vertical dashed lines the points of the spectrum where kinks or
discontinuities can happen. For the moment it suffices to say that both the minimal and maximal values of those
two event shapes, $e_{\rm min}$ and $e_{\rm max}$, depend on the reduced mass and become equal for
$\hat m = 1/2$. The discussion on how these points appear and their analytic expressions shall be presented
elsewhere. It is worth commenting on the result for heavy jet mass, since its behavior is different
depending on the value of $\hat m$. There is a critical value
$\hat m_{\rm crit} \approx 0.286169$ above which there is a region with zero cross
section within the physical spectrum. The cumulative cross section is totally flat in this patch. This
behavior is of kinematic origin and therefore takes place for both currents. There is no such effect for
$2$-jettiness.

\begin{figure*}[t!]
\includegraphics[width=0.46\textwidth]{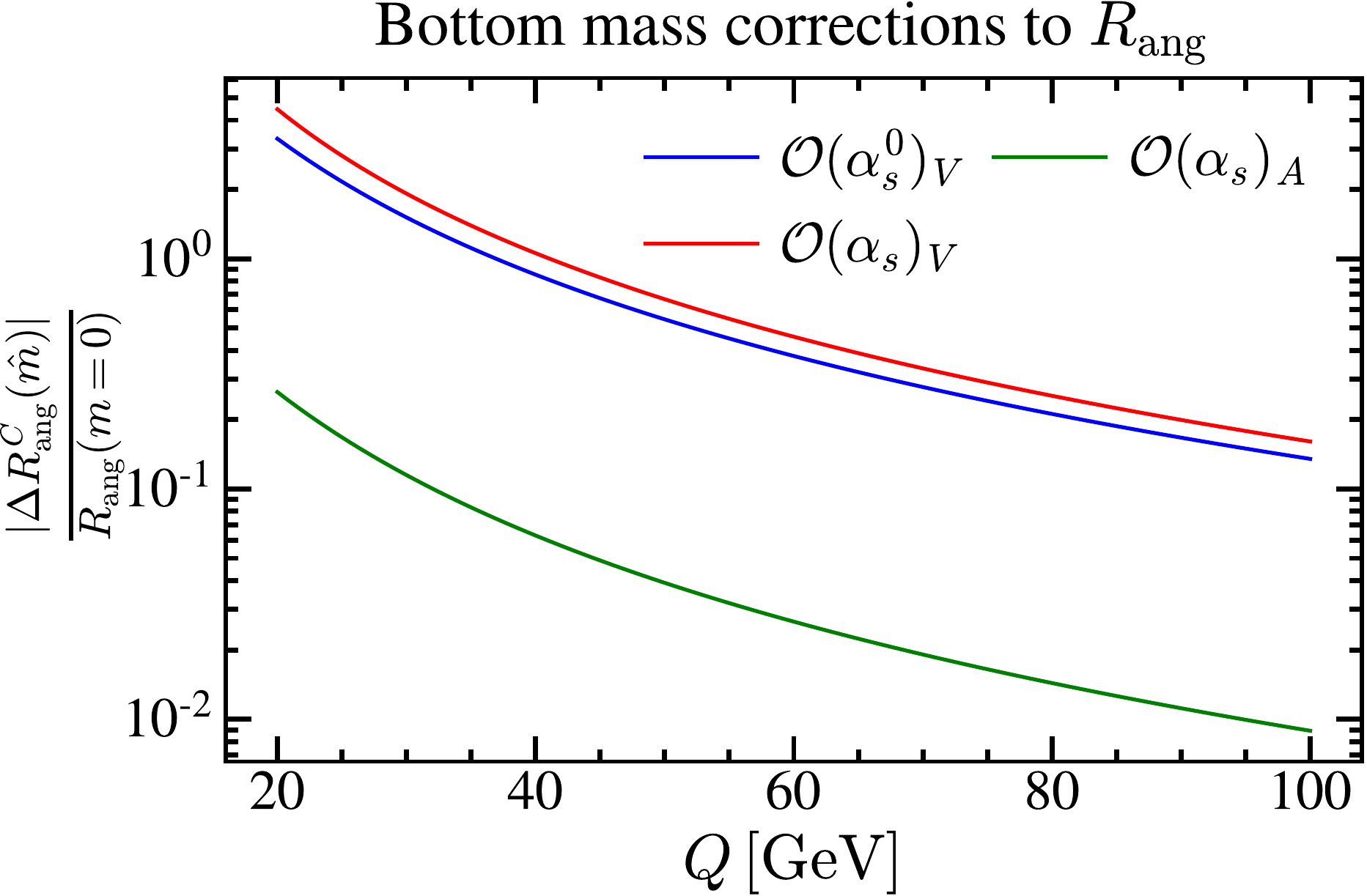}
\includegraphics[width=0.46\textwidth]{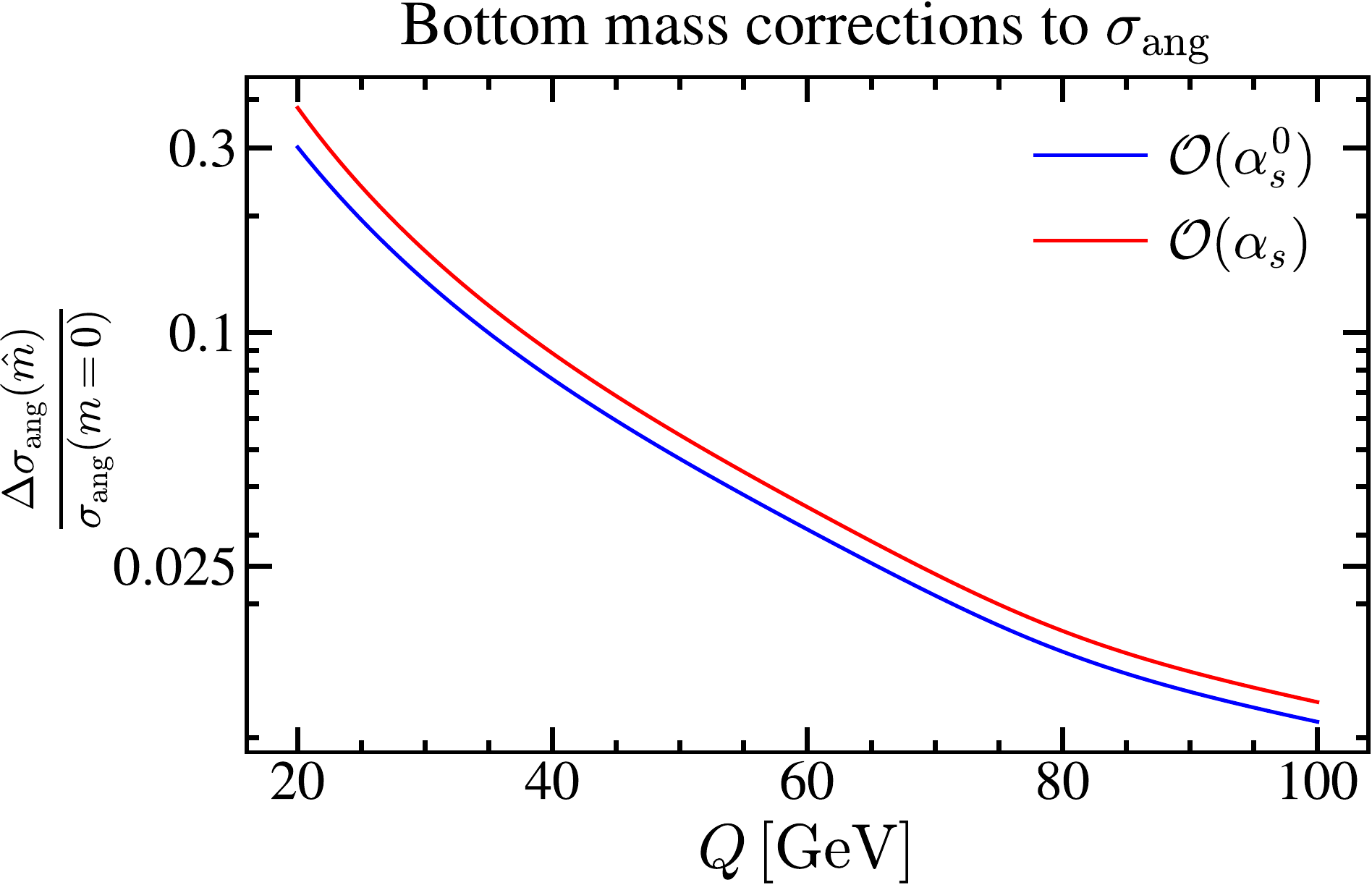}
\caption{Bottom mass corrections to the total angular cross section.}
\label{fig:bottom}
\end{figure*}

Finally, we present an analysis of the impact of the bottom quark mass as a correction to the massless prediction for the
total oriented cross section $R_{\rm ang}$, see Fig.~\ref{fig:bottom}. This observable is of high relevance since it is a prominent
candidate to determine the strong coupling fitting to experimental data. This is so because, due to
inclusive nature, it suffers from small hadronization effects and Sudakov-log resummation might not be
necessary, but at the same time, at lowest order it is already linearly dependent on $\alpha_s$. In our analysis, first we assume
one can experimentally tag on bottom quarks (for example through the presence of $B$ mesons in the
final state) and on the current (for instance counting the number of pions), and second, consider a more
realistic scenario in which no tagging is applied. In this latter situation we find that
the massive correction at NLO for $Q=20, 50$ and $100\,$GeV is as large as $38\%$, $4.27\%$ and
$1.11\%$, respectively. Therefore it cannot be neglected in any analysis that aims for high precision.
\section{Conclusions}
We have presented fixed-order results for oriented event shapes at NLO considering massive quarks. The computation
has been organized such that the relevant angular distribution is projected out at a very early
stage, and chosen a $d$-dimensional normalization that avoids having unphysical logs at intermediate
steps. Bottom mass corrections turn out to be a very large correction to the massless approximation for LEP and
smaller energies. An immediate application for our computation is the determination of the strong couplings
by a comparison to existing experimental data on $R_{\rm ang}$.
Our results for the total oriented cross section reveal a Sommerfeld enhancement at threshold,
such that this observable might be useful to determine the top quark mass at a future $e^+e^-$ collider
through threshold scans.

\bibliography{../JHEP/thrust3}

\end{document}